\documentclass[reprint,aps,amsmath,amssymb,superscriptaddress,floatfix]{revtex4-2}

\usepackage[utf8]{inputenc}
\usepackage{setspace}
\usepackage{graphicx}
\usepackage{color}
\usepackage{soul}
\usepackage{hyperref}
\usepackage{miller}
\usepackage{pdfpages}
\usepackage{bbold}
\usepackage{xspace}
\usepackage{natbib}
\usepackage{natmove}
\setcitestyle{super}

\newcommand{\simlt}
      {\ifmmode       { \raisebox{-.8em}{$<$}\atop\sim}
         \else        {$\raisebox{-.8em}{$<$}\atop\sim$}
      \fi}

\newcommand{\Tc}{\ensuremath{T_\mathrm{C}}\xspace}
\newcommand{\TN}{\ensuremath{T_\mathrm{N}}\xspace}
\newcommand{\ea}{{\it et al.}\xspace}

\makeatletter
\def\@hangfrom@section#1#2#3{\@hangfrom{#1#2}#3}
\def\@hangfroms@section#1#2{#1#2}
\makeatother

\makeatletter
\AtBeginDocument{\let\LS@rot\@undefined}
\makeatother

\usepackage[version=4]{mhchem}

\begin{document}


\title{From ferromagnetic semiconductor to anti-ferromagnetic metal \\in epitaxial Cr$_x$Te$_y$ monolayers}

\author{Naina Kushwaha}
\affiliation{SUPA, School of Physics and Astronomy, University of St Andrews, North Haugh, St Andrews, KY16 9SS, United Kingdom}
\affiliation{STFC Central Laser Facility, Research Complex at Harwell, Harwell Campus, Didcot OX11 0QX, United Kingdom}
\author{Olivia Armitage}
\affiliation{SUPA, School of Physics and Astronomy, University of St Andrews, North Haugh, St Andrews, KY16 9SS, United Kingdom}
\author{Brendan Edwards}
\affiliation{SUPA, School of Physics and Astronomy, University of St Andrews, North Haugh, St Andrews, KY16 9SS, United Kingdom}

\author{Liam Trzaska}
\affiliation{SUPA, School of Physics and Astronomy, University of St Andrews, North Haugh, St Andrews, KY16 9SS, United Kingdom}

\author{Peter Bencok}
\author{Gerrit van der Laan}
\affiliation{Diamond Light Source, Harwell Science and Innovation Campus, Didcot, OX11 ODE, United Kingdom}
\author{Peter Wahl}
\email[Correspondence to: ]{wahl@st-andrews.ac.uk.}
\affiliation{SUPA, School of Physics and Astronomy, University of St Andrews, North Haugh, St Andrews, KY16 9SS, United Kingdom}
\affiliation{Physikalisches Institut, Universität Bonn, Nussallee 12, 53115 Bonn, Germany}
\author{Phil D. C. King}
\email[Correspondence to: ]{pdk6@st-andrews.ac.uk.}
\affiliation{SUPA, School of Physics and Astronomy, University of St Andrews, North Haugh, St Andrews, KY16 9SS, United Kingdom}
\author{Akhil Rajan}
\affiliation{SUPA, School of Physics and Astronomy, University of St Andrews, North Haugh, St Andrews, KY16 9SS, United Kingdom}

\date{\today}

\begin{abstract}
 Chromium ditelluride, CrTe$_2$, is an attractive candidate van der Waals material for hosting 2D magnetism. However, how the room-temperature ferromagnetism of the bulk evolves as the sample is thinned to the single-layer limit has proved controversial. This, in part, reflects its metastable nature, vs.\ a series of more stable self-intercalation compounds with higher relative Cr:Te stoichiometry. Here, exploiting a recently-developed method for enhancing nucleation in molecular beam epitaxy growth of transition-metal chalcogenides, we demonstrate the selective stabilisation of high-coverage CrTe$_2$ and Cr$_{2+\varepsilon}$Te$_3$ epitaxial monolayers. Combining X-ray magnetic circular dichroism, scanning tunnelling microscopy, and temperature-dependent angle-resolved photoemission, we demonstrate that both compounds order magnetically with a similar \Tc. We find, however, that monolayer CrTe$_2$ forms as an anti-ferromagnetic metal, while monolayer Cr$_{2+\varepsilon}$Te$_3$ hosts an intrinsic ferromagnetic semiconducting state. This work thus demonstrates that control over the self-intercalation of metastable Cr-based chalcogenides provides a powerful route for tuning both their metallicity and magnetic structure, establishing the Cr-Te system as a flexible materials class for future 2D spintronics.
\end{abstract}


\maketitle

\section{Introduction}

The MX$_2$ transition metal dichalcogenides (TMDs, M=transition metal, X=\{S,Se,Te\}) are one of the most prominent current classes of 2D materials \cite{manzeli20172d,schneider2018two,chen2023excitonic}. However, despite the wide array of crystal structures and transition metal $d$-electron occupancies which can be realised across this family,~\cite{Chhowalla2013} most TMDs are non-magnetic. In contrast, magnetic order has been more notable in the layered halides such as CrI$_3$, or ternary chalcogenides such as CrGeTe$_3$ or Fe$_3$GeTe$_2$, where long-range magnetic order has been found to persist to the bi- or even monolayer (ML) limit \cite{huang2017layer,gong2017discovery,fei2018two}. Nonetheless, for materials simplicity, as well as structural and fabrication compatibility for the formation of integrated heterostructures, it remains of strong interest to develop intrinsic 2D magnetic systems within the TMD family. 

One of the most promising candidates is CrTe$_2$. While this compound is only metastable, Freitas {\it et al.} \cite{freitas2015ferromagnetism} successfully demonstrated the synthesis of bulk 1T-CrTe$_2$ samples by de-intercalating K from KCrTe$_2$. The de-intercalated samples exhibit itinerant ferromagnetism with a Curie temperature (\Tc) of 310 K. This suggests a robust ferromagnetic order in this quasi-2D layered system, raising hopes to maintain long-range order down to the single-layer thickness limit. However, studies of the evolution of its magnetic order with sample thickness have proved controversial to date. Lingjia \ea \cite{meng2021anomalous} fabricated multi-layer 1T-CrTe$_2$ by chemical vapor deposition (CVD), finding an unconventional increase in the reported Curie temperature with decreasing film thickness down to $\approx8$~nm, but with a \Tc in 130~nm-thick samples only around half that of the bulk crystals. Combined first-principles and Monte Carlo calculations have suggested that ferromagnetism should be stable in the ML limit \cite{yao2023control}, albeit with a strongly reduced \Tc, and with a significant tunability with applied strain, electron and hole doping, and varying the number of layers. There have been several attempts to study this ML regime in samples grown by molecular-beam epitaxy (MBE). Zhang~\ea \cite{zhang2021room} reported the observation of a persistent ferromagnetic order, with \Tc above room temperature in few-layer films, and dropping to 200~K in the monolayer. However, the samples investigated had an anomalously large in-plane lattice constant. In contrast, from spin-polarised scanning tunnelling microscopy experiments, Xian \ea \cite{xian2022spin} concluded that ML-CrTe$_2$ orders antiferromagnetically, although with an unknown N\'eel temperature, \TN. Further studies of the putative magnetic order of ML-CrTe$_2$ are thus urgently required.

 \begin{figure*}
   \centering
    \includegraphics[width=\linewidth]{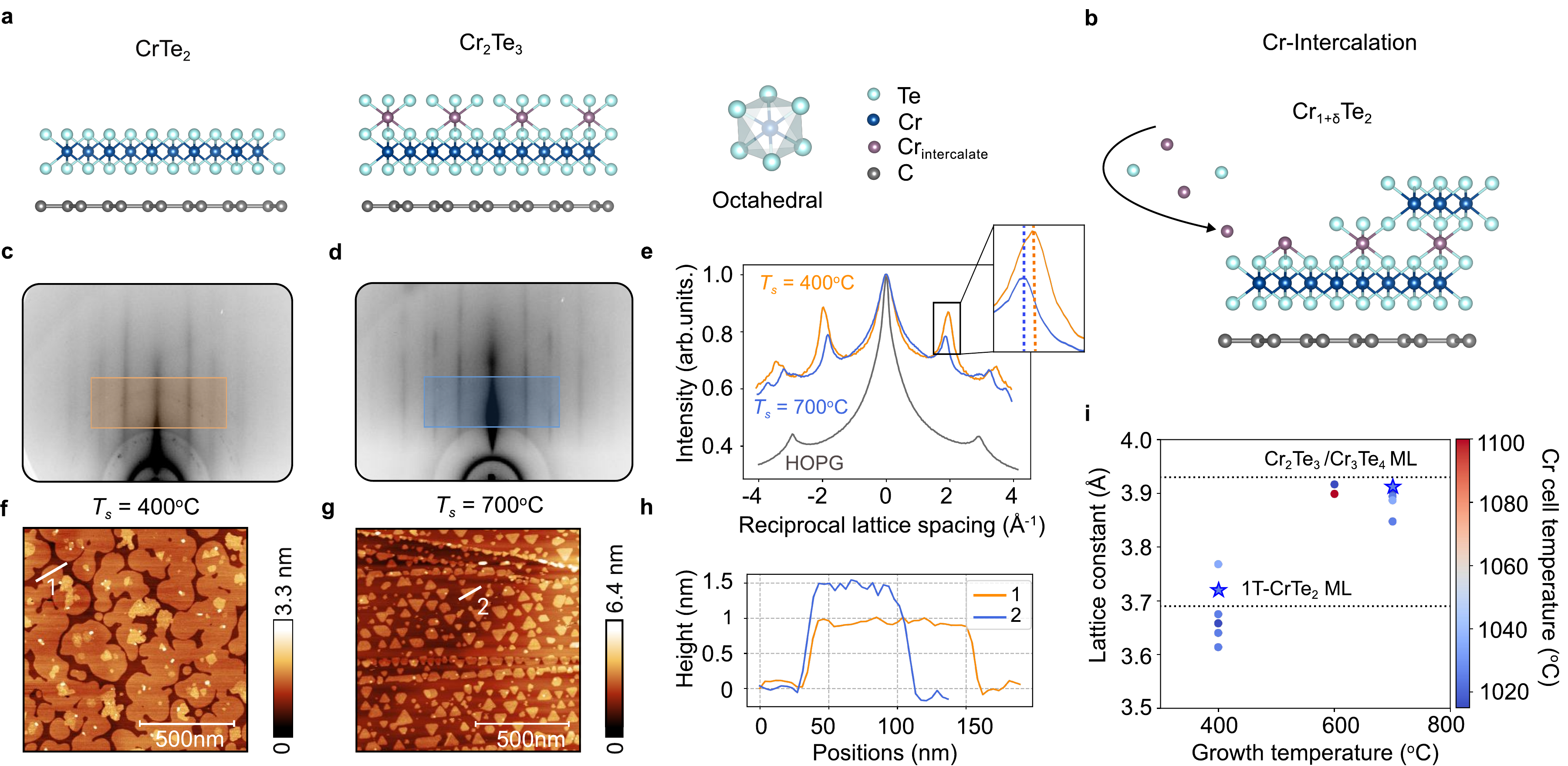}
    \caption{{\bf Selective epitaxial growth of Cr$_x$Te$_y$ .} (a) Side view of the crystal structure of ML-CrTe$_2$ and ML-Cr$_2$Te$_3$ on an HOPG substrate. (b) Schematic of the formation of the Cr$_{1+\delta}$Te$_2$ phases with Cr atoms self-intercalating within the van der Waals gap of CrTe$_2$. (c,d) RHEED patterns of samples grown at a substrate temperature of (c) $T_s=400^\circ$C and (d) $T_s=700^\circ$C [Cr cell temperature: $1025^\circ$C], and (e) corresponding intensity profiles extracted from the shaded regions in (c,d). The extracted in-plane lattice constants (stars in (i)) are consistent with CrTe$_2$ and [Cr$_2$Te$_3$, Cr$_3$Te$_4$], respectively. (f,g) AFM images for the samples shown in (c,d). (h) AFM line profiles corresponding to the lines shown in (f,g), revealing the different thickness of the monolayer islands, again consistent with CrTe$_2$ and [Cr$_2$Te$_3$, Cr$_3$Te$_4$], respectively. (i) Obtained growth phase diagram of Cr$_x$Te$_y$ as a function of substrate and Cr effusion cell temperature, as determined from lattice constant measurements from RHEED.}
    \label{fig:growth_temp_control}
\end{figure*}

A challenge is the metastable nature of this compound. In an ionic picture, the Cr ions in CrTe$_2$ would be expected in a nominal 4+ configuration. This would lead to an unstable $d^2$ charge count. The compound would thus be expected to preferentially form in a bilayer configuration, with additional Cr atoms intercalating within the van der Waals gap to move towards the stable Cr $d^3$ (Cr$^{3+}$) configuration (Fig.~\ref{fig:growth_temp_control}(a,b)). Indeed, a variety of intercalated Cr$_{1+\delta}$Te$_2$ phases are known to exist (Cr$_2$Te$_3$, Cr$_3$Te$_4$, Cr$_5$Te$_8$), depending on the partial occupancy of intercalated Cr atoms. Lasek~\ea~\cite{lasek2020molecular,lasek2022van} have previously demonstrated a strong sensitivity of the stabilised Cr-Te phase to the growth conditions used as well as post-growth annealing, with a marked propensity for transformation to the self-intercalated phases, and only small monolayer patches of CrTe$_2$ obtained. The more Cr-rich intercalated phases have been shown to exhibit clear magnetic signatures \cite{chaluvadi2024uncovering, Zhong2023,chua2021room,chen2022air}, making it challenging to isolate and characterise the intrinsic magnetic properties of ML-CrTe$_2$. 

In this work, we present a controlled methodology for realising the MBE growth of different Cr$_x$Te$_y$ phases. Utilising an ion-assisted method~\cite{advmater2024} to enhance the nucleation of the growing film, we realise much higher coverage of nearly phase-pure CrTe$_2$ and Cr$_{2+\varepsilon}$Te$_3$ monolayers, allowing their subsequent investigation by core-hole, photoemission, and scanning-probe spectroscopies. Through this, we identify that both compounds order magnetically in the monolayer limit, but with markedly distinct magnetic and electronic structures.

\begin{figure*}
   \centering
    \includegraphics[width=0.8\linewidth]{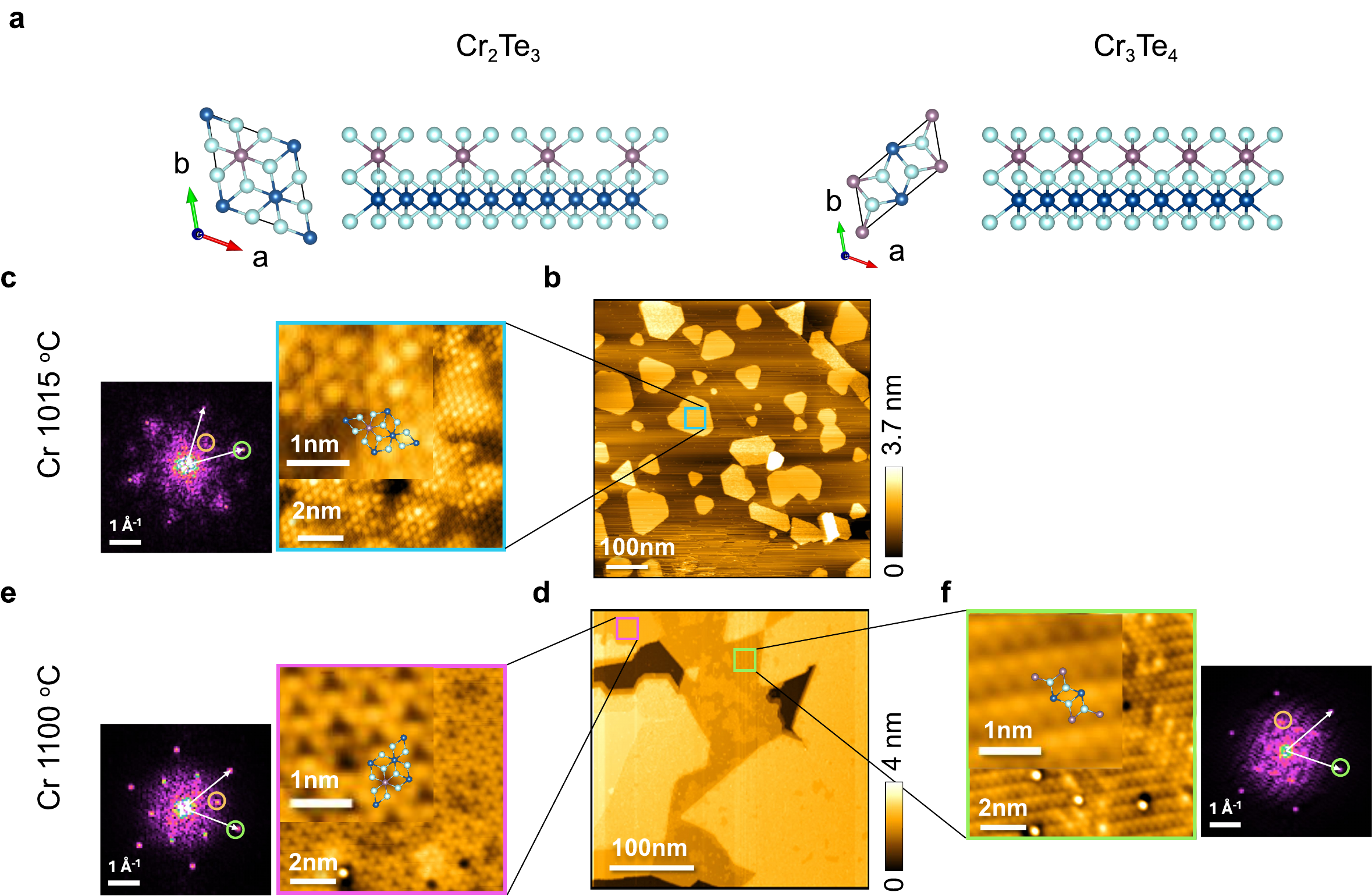}
     \caption{{\bf Stoichiometry control of Cr-intercalated phases.} (a) Schematic top view (projected along $[0001]$) and side view (projected along $[2\overline{1}\overline{1}0]$ and $[10\overline{1}0]$ respectively) of the crystal structure of a Cr$_2$Te$_3$ and Cr$_3$Te$_4$ ML. The self-intercalated Cr atoms are shown in pink. (b-f) STM images of Cr$_x$Te$_y$ MLs synthesized using a Cr effusion cell temperature of (b,c) $1015^\circ$C and (d-f) $1100^\circ$C: (b,d) overview measurements; (c,e,f) high resolution measurements from the regions shown in (b,d), with corresponding magnified views (inset) and Fourier transforms shown. The Fourier images show Bragg peaks corresponding to the CrTe$_2$ lattice (green circles). In addition, the regions in (c) and (e) show an additional $(\sqrt{3}\times\sqrt{3})R30^{\circ}$ superstructure order, corresponding to Cr$_2$Te$_3$, while (f) shows a $(2\times1)$ superstructure, consistent with Cr$_3$Te$_4$ (highlighted by orange circles in the FFT). Tunnelling setpoints: (b) $V_\mathrm{s}$ = 900 mV, $I_\mathrm{s}$ = 50 pA,(c) $V_\mathrm{s}$ = -300 mV, $I_\mathrm{s}$ = 75 pA, (d) $V_\mathrm{s}$ = 900 mV, $I_\mathrm{s}$ = 60 pA, (e) $V_\mathrm{s}$ = -200 mV, $I_\mathrm{s}$ = 100 pA, (f) $V_\mathrm{s}$ = -200 mV, $I_\mathrm{s}$ = 100 pA.}
     \label{fig:Cr_rich_vs_poor}
\end{figure*}

\section{Results}
\subsection*{Epitaxial growth of phase-selective Cr$_x$Te$_y$}
 Figure~\ref{fig:growth_temp_control} shows the growth of monolayer Cr$_x$Te$_y$ epitaxial layers via MBE, with control over the stoichiometry obtained by tuning the growth conditions used. We show in Fig.~\ref{fig:growth_temp_control}(c-d, f-g) reflection high-energy electron diffraction (RHEED) and atomic force microscopy (AFM) measurements from samples grown on highly oriented pyrolytic graphite (HOPG) substrates (see Methods), at a substrate temperature of $T_s=400^\circ$C and $T_s=700^\circ$C, respectively. Clear diffraction streaks from the growing epilayer as well as the underlying substrate are evident. Comparing the intensity profiles extracted from these RHEED images (Fig.~\ref{fig:growth_temp_control}(e)), it is clear that the diffraction streaks corresponding to the epilayer are located at larger wavevector (corresponding to a smaller in-plane lattice constant) for the sample grown at the lower substrate temperature. From fitting the extracted line profiles across multiple grown samples, we find lattice constants of $a=3.66\pm0.04$~\AA{} ($a=3.88\pm0.02$~\AA) for the samples grown at $T_s=400^\circ$C ($700^\circ$C), respectively. These values are within experimental error of the in-plane lattice constant expected for CrTe$_2$ and the Cr-intercalated (Cr$_2$Te$_3$, Cr$_3$Te$_4$, Cr$_5$Te$_8$) phases, respectively~\cite{lasek2020molecular}. 

 Consistent with this, our AFM measurements (Fig.~\ref{fig:growth_temp_control}(f-h) indicate a clear impact of substrate temperature on the as-grown epi-islands. At $T_s=400^\circ$C, while limited adatom mobility restricts edge diffusion, resulting in irregular island shapes, the substrate is nearly entirely covered with a single epilayer, with just occasional bilayer patches visible, and a few gaps to the substrate below. Conversely, at the higher substrate temperature, enhanced adatom surface diffusion leads to more triangular island shapes~\cite{Rajan2020}, but a lower surface coverage indicating a slower growth rate. Extracted height profiles from the first epilayer to the substrate below (Fig.~\ref{fig:growth_temp_control}(h)) yield ML step heights of $\sim 1 $nm and $\sim 1.5$ nm for the samples grown at $T_s=400^\circ$C and $700^\circ$C, respectively. The former is consistent with the AFM step height typically obtained for the first monolayer of an MX$_2$ TMD, again indicating that our samples grown at $T_s=400^\circ$C are CrTe$_2$ monolayers. In contrast, the higher minimum step height observed for the higher temperature growths again point to the fact that growth proceeds immediately to the formation of a self-intercalated phase under these conditions. We note that the small additional height over the CrTe$_2$ monolayer, as well as the small step to the bilayer patches visible in Fig.~\ref{fig:growth_temp_control}(f) (see also Supplementary Fig.~1), indicates growth of a quintuple layer Te-Cr-Te-Cr-Te structure as shown in Fig.~\ref{fig:growth_temp_control}(a), rather than the fabrication of a full bilayer of CrTe$_2$ with additional Cr self-intercalation (Fig.~\ref{fig:growth_temp_control}(b)). We refer to the former as a monolayer of the self-intercalated structure in the following.  

 \begin{figure}
   \centering
    \includegraphics[width=\columnwidth]{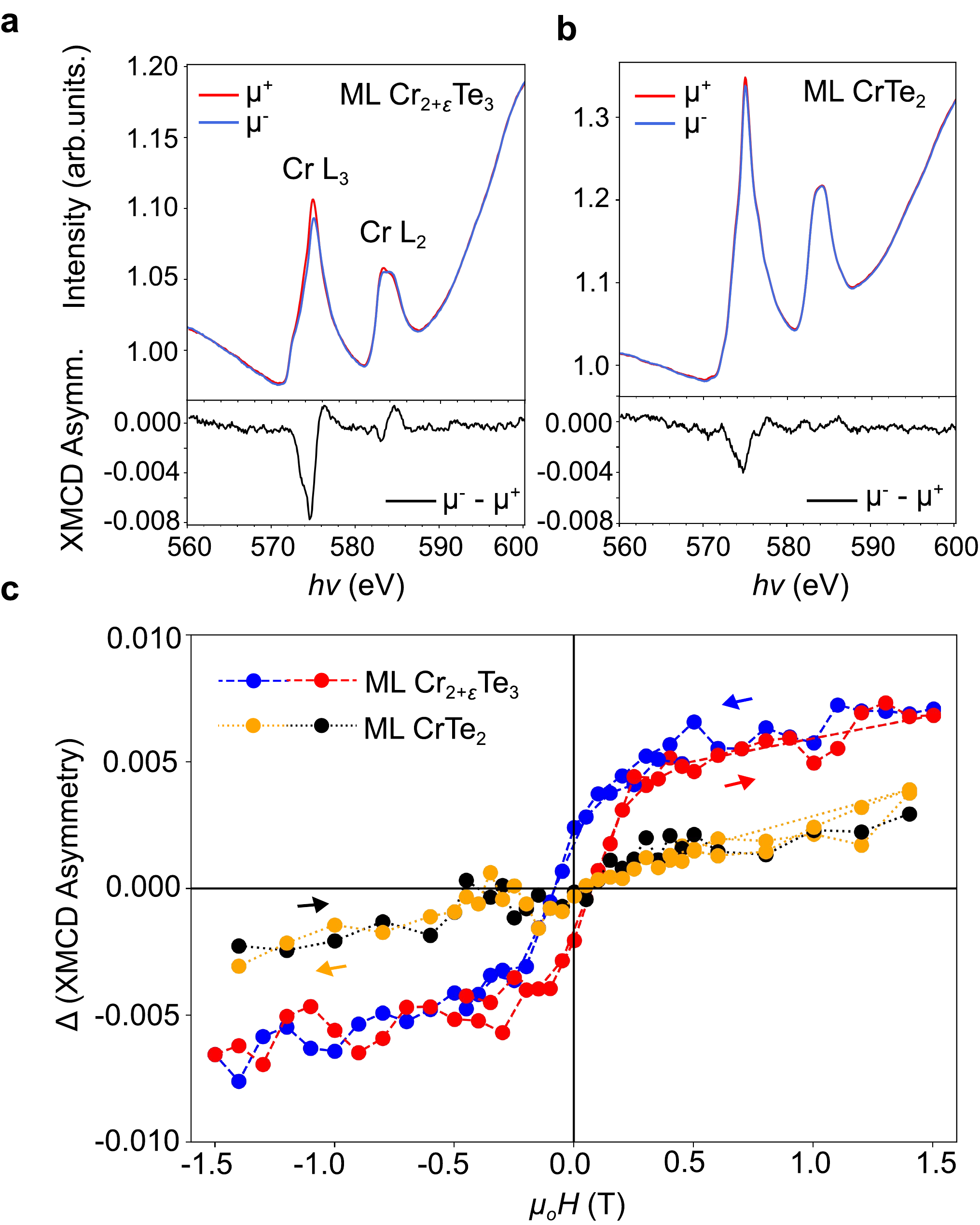}
    \caption{{\bf X-ray magnetic circular dichroism of Cr$_{2+\varepsilon}$Te$_3$ and CrTe$_2$.} (a,b) Total electron yield XAS measurements performed using left- and right-circularly polarised X-rays ($\mu^\pm$), measured from (a) Cr$_{2+\varepsilon}$Te$_3$ and (b) CrTe$_2$ monolayer samples across the Cr $L_{2,3}$ absorption edge at 10~K in a 1~T applied magnetic field. The corresponding XMCD asymmetry [$(I^{\mu^-}-I^{\mu^+})/(I^{\mu^-}+I^{\mu^+})$] is shown below. (c) Peak-to-peak span of the XMCD $L_{3}$-edge asymmetry ($\Delta$(XMCD Asymm.), see Supplementary Fig.~3) as a function of the applied magnetic field. Arrows indicate the direction of the magnetic field ramping.}
    \label{fig:XMCD}
     \end{figure}
     
The above findings are broadly in line with those reported by Lasek~\ea~\cite{lasek2020molecular}, but we are able to stabilise the CrTe$_2$ phase at higher substrate temperatures and with much larger island sizes across the substrate. This is because our samples are grown using co-evaporation of a minute quantity of Ge from an electron-beam evaporator (see Methods). We have recently shown how the exposure of the substrate to the flux of excited Ge ions from such a source induces substrate defects which dramatically enhance the nucleation of the growing epilayer \cite{advmater2024}. Indeed, under the same growth conditions used here but without co-exposure to the Ge ions, we find significantly more limited coverage of CrTe$_2$ even with an extended growth duration, with growth occurring primarily along the substrate step edges (see Supplementary Fig.~2). The use of our ion-assisted nucleation strategy here is thus essential to obtain larger-area and significantly more uniform growth of ML-CrTe$_2$, and leads to a significantly enhanced growth window. Nonetheless, our measurements in Fig.~\ref{fig:growth_temp_control}(i) indicate a gradual increase in the lattice constant with increasing Cr flux during the growth, pointing to the onset of self-intercalation and bilayer formation, and indicating the importance of controlling both the growth temperature and the Cr flux for stabilising true ML-CrTe$_2$.

While the obtained lattice constants and AFM step heights allow the ready identification of CrTe$_2$, they do not allow a reliable determination of the stoichiometry of the self-intercalated phases grown at higher temperature, which all host similar in-plane lattice constants and ML-layer heights to each other. We thus adopt atomic-scale imaging using scanning tunnelling microscopy, which is sensitive to the different lateral periodicities of the self-intercalated Cr ions for the different stoichiometries~\cite{lasek2020molecular}. As shown in Fig.~\ref{fig:Cr_rich_vs_poor}(b,c), for a sample grown with a substrate temperature of $T_s=600^\circ$C and a low Cr cell temperature of $1015^\circ$C, a set of islands with largely uniform heights are found distributed across the sample. Atomic-resolution imaging (Fig.~\ref{fig:Cr_rich_vs_poor}(c)) shows a disordered lattice, but with an average additional super-periodicity of $(\sqrt{3}\times\sqrt{3})R30^{\circ}$ evident in fast Fourier transform (FFT) analysis. This is consistent with the additional periodicity expected for intercalated Cr atoms in the van der Waals gap for a Cr$_2$Te$_3$ phase (Fig.~\ref{fig:Cr_rich_vs_poor}(a)). Similar periodicity is found at any island investigated, allowing us to identify the dominant phase here as Cr$_2$Te$_3$, although with some site disorder evident. We thus refer to this phase as Cr$_{2+\varepsilon}$Te$_3$. In contrast, when using a higher Cr cell temperature of $1100^\circ$C  during the growth (Fig.~\ref{fig:Cr_rich_vs_poor}(d-f)), we find higher multi-layer coverage and much more pronounced variations across the sample surface. Some regions (Fig.~\ref{fig:Cr_rich_vs_poor}(e)) again show a $(\sqrt{3}\times\sqrt{3})R30^{\circ}$ order, but with greater local-scale order and uniformity than for the sample grown with lower Cr flux. In contrast, other regions on the same sample show a $(2\times1)$ superstructure (Fig.~\ref{fig:Cr_rich_vs_poor}(f)), indicative of Cr$_3$Te$_4$. This points to an intrinsic phase separation into more Cr-rich and Cr-poor regions which - while it may lead to greater crystalline order on a local scale - leads to undesired mesoscopic variations in the stoichiometry. For the remainder of this work, we thus restrict our studies to the low Cr flux regime, where Cr$_{2+\varepsilon}$Te$_3$ is found across the sample surface.

  \begin{figure*}
   \centering
    \includegraphics[scale=0.27]{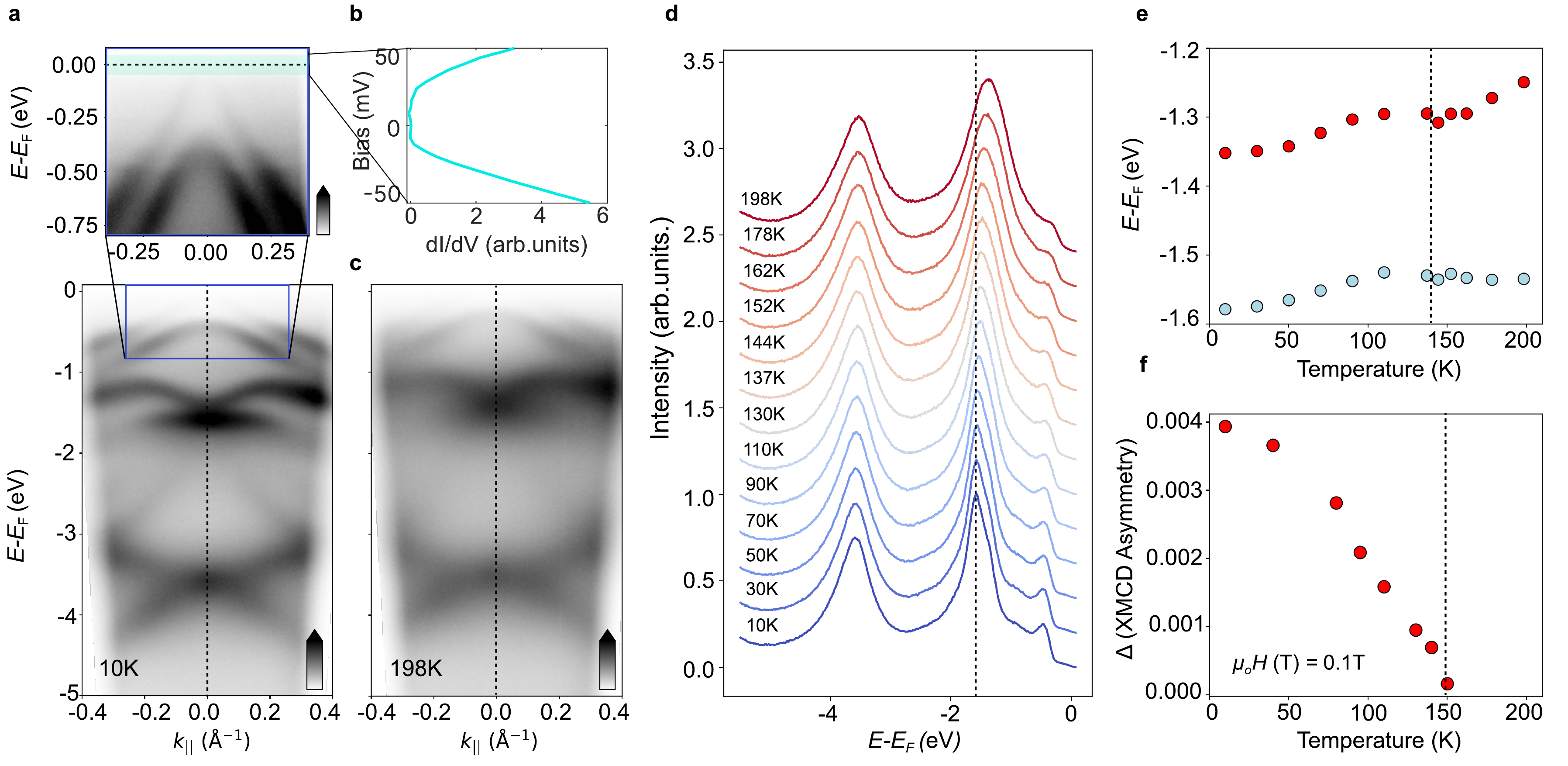}
    \caption{{\bf Electronic structure of ML-Cr$_{2+\varepsilon}$Te$_3$.} (a) ARPES measurements of a 1 ML Cr$_{2+\varepsilon}$Te$_3$ sample measured at $T=10$~K. A magnified view in the vicinity of the Fermi level is shown with enhanced contrast inset, revealing the top of a pair of hole-like valence bands in close proximity to $E_\mathrm{F}$. (b) Measured $dI/dV$ spectra from an equivalent sample in scanning tunnelling spectroscopy ($V_\mathrm{s}$ = 120 mV, $I_\mathrm{s}$ = 400 pA, lock-in amplitude 3 mV). A small semiconducting gap is found around the Fermi level. (c) ARPES measurements at $T=198$~K and (d) EDCs extracted at the $\Gamma$ point obtained from temperature-dependent measurements between 10 and 198~K. A small but clear band shift of several features around a binding energy of 1.5~eV are visible (dashed line), and extracted from peak fitting analysis in (e). (f) Cr $L_{3}$-edge $\Delta$(XMCD Assym.) as a function of temperature, measured in an applied field of 0.1~T. The loss of the dichroism signal in the measured XMCD co-incides with an anomaly in temperature-dependent band shifts from ARPES (dashed lines in (e,f)), indicating both are linked to the onset of magnetic order.}
    \label{fig:Cr2Te3_ARPES}
   \end{figure*}
\subsection*{Dichotomy in magnetic order and electronic structure}
To investigate the intrinsic magnetic order of such Cr$_{2+\varepsilon}$Te$_3$ monolayers, as well as our CrTe$_2$ samples, we employed X-ray Absorption Spectroscopy (XAS) and X-ray Magnetic Circular Dichroism (XMCD). Performing these measurements at the Cr $L_{2,3}$ absorption edge (Fig.~\ref{fig:XMCD}(a,b)) provides a sensitive, element-specific probe of magnetic order in the monolayer samples. In an applied magnetic field of 1~T, our Cr$_{2+\varepsilon}$Te$_3$ monolayers exhibit a distinctly resolvable magnetic dichroism (Fig.~\ref{fig:XMCD}(a)), consistent with that recently reported for other Cr$_2$Te$_3$ monolayers~\cite{Zhong2023}. Its form is similar to the XMCD observed in other nominally Cr$^{3+}$ 2D magnet families such as CrGeTe$_3$, CrSiTe$_3$ and CrI$_3$~\cite{Watson2020, achinuq2022covalent,huang2018electrical}, pointing to a dominant spin moment and small orbital moment. Tracking the peak asymmetry of the measured XMCD at the Cr $L_{3}$-edge as a function of applied magnetic field (Fig.~\ref{fig:XMCD}(c)), we find a clear hysteretic behaviour with a saturation field of $\sim$~0.5~T, and a coercive field of $\sim$~0.1~T. This unambiguously confirms that ML-Cr$_{2+\varepsilon}$Te$_3$ exhibits long-range ferromagnetic order. In contrast, the XMCD signal that we observe for ML-CrTe$_2$ (Fig.~\ref{fig:XMCD}(b)) is both weaker than for its self-intercalated cousin, and it displays an almost linear dependence on the applied field (Fig.~\ref{fig:XMCD}(c)). A complete lack of any magnetic hysteresis rules out that CrTe$_2$ has a ferromagnetic order in the monolayer limit, in contrast to prior reports~\cite{zhang2021room}. Nonetheless, a clearly resolvable dichroic signal in applied field still points to the formation of a well-developed local moment in ML-CrTe$_2$.

To further explore these contrasting magnetic signatures, we performed additional spectroscopic characterisation of our samples. Fig.~\ref{fig:Cr2Te3_ARPES}(a) shows the measured low-temperature electronic structure of our Cr$_{2+\varepsilon}$Te$_3$ monolayers from angle-resolved photoemission spectroscopy (ARPES). Consistent with previous reports~\cite{Zhong2023}, we find a pair of dispersive states in the vicinity of the Fermi level, which can be assigned as the spin-orbit split Te~5$p$ states.~\cite{Watson2020,Antonelli2022} Somewhat flatter states of expected Cr character are obtained at higher energies of $E-E_{\mathrm{F}}\approx\!1.5$~eV. This is consistent with a nominal Cr$^{3+}$ charge state, and hence a nominally half-filled spin-polarised Cr $t_{2g}$ band being pushed below the Fermi level by the exchange splitting with the formation of a magnetic moment. Nonetheless, we note that this Cr-derived state is still rather dispersive in our measurements, pointing to a significant Cr-Te hybridisation. This is qualitatively consistent with our XAS measurements (Fig.~\ref{fig:XMCD}(a)) which are rather smeared out, without a significant pre-peak structure, thus suggesting a significant $d$-$p$ admixture. The band top of the Te $p$-derived states approach, but do not quite reach, the Fermi level (Fig.~\ref{fig:Cr2Te3_ARPES}(a)) while our corresponding spectroscopy measurements performed with low-temperature STM (Fig.~\ref{fig:Cr2Te3_ARPES}(b)) confirm a gapped state around the Fermi level (see also corresponding quasiparticle interference measurements in Supplementary Fig.~4). We note that previous measurements~\cite{Zhong2023} found the Te-derived states to cross the Fermi level, leading to a metallic state. We attribute this discrepancy to a likely role of Cr off-stoichiometry from the nominal Cr$_2$Te$_3$ composition. Here, however, our combined tunnelling and photoemission measurements allow us to unambiguously assign the ground state of ML-Cr$_2$Te$_3$ as a narrow-gap semiconductor. This rules out any Stoner character of the observed magnetic order in this system, supporting recent conclusions that the magnetic order in the monolayer limit of Cr$_2$Te$_3$ is of local-moment character.~\cite{Zhong2023}

\begin{figure*}
   \centering
    \includegraphics[scale=0.3]{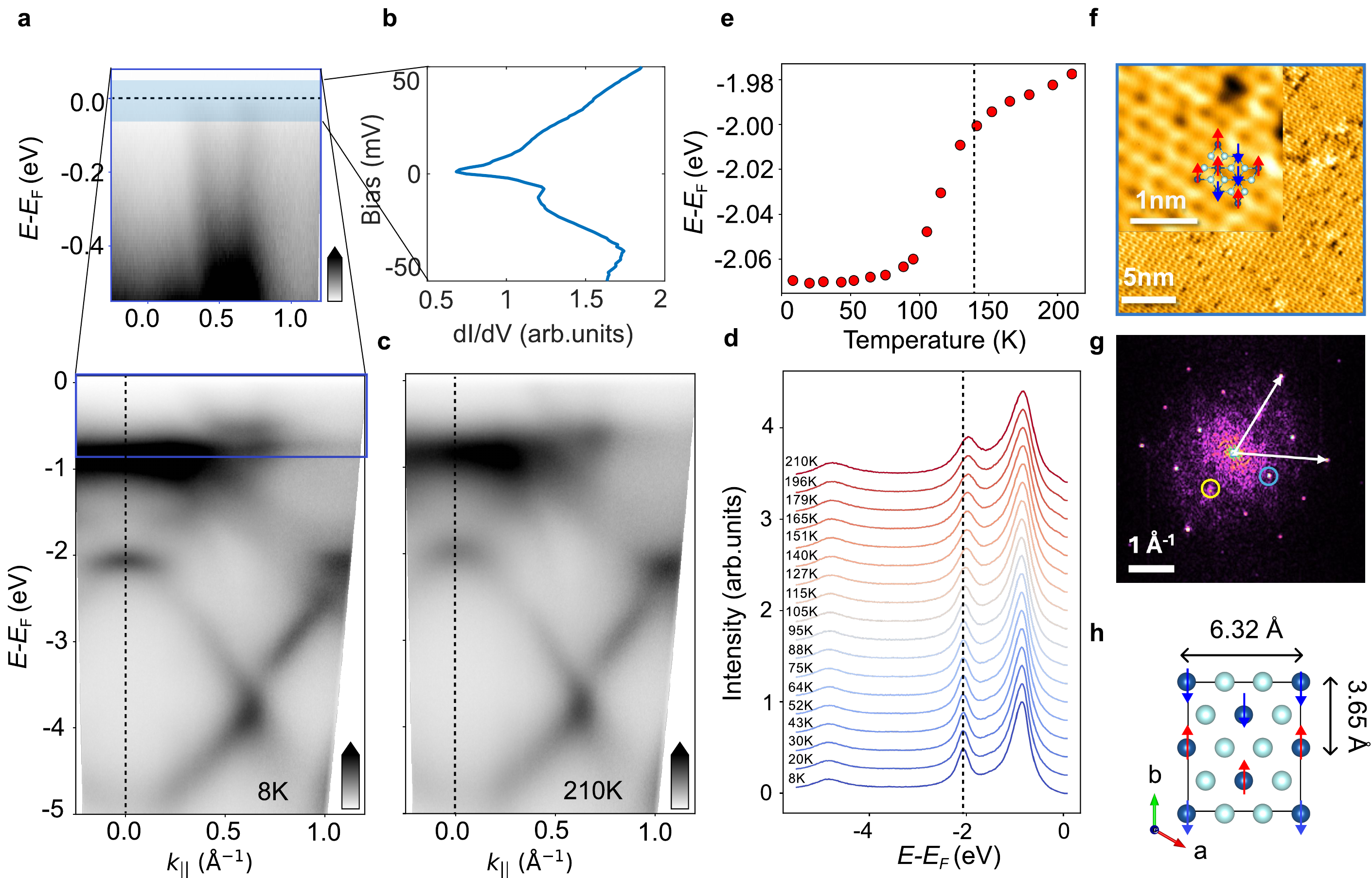}
     \caption{{\bf Antiferromagnetic order in ML-CrTe$_2$.} (a) ARPES measurements of a 1 ML CrTe$_2$ sample measured at $T=8$~K. The inset shows metallic states crossing the Fermi level, consistent with (b) measured $dI/dV$ spectra from scanning tunnelling spectroscopy ($V_\mathrm{s}$ = 50 mV, $I_\mathrm{s}$ = 100 pA, lock-in amplitude 800 $\mu$V). (c) ARPES measurements at $T=210$~K, and (d) corresponding temperature-dependent EDCs extracted at the $\Gamma$ point, showing (e) a clear temperature-dependent anomaly at $T\approx140$~K. (f,g) Atomically resolved STM measurements (f) and associated FFT analysis (g) of ML-CrTe$_2$ ($V_\mathrm{s}$ = 250 mV, $I_\mathrm{s}$ = 250 pA), showing an underlying structural periodicity associated with a $(1\times2)$ modulation (yellow circle) with an additional real-space zigzag pattern associated with a $(2\times1)$ supermodulation (blue circle), respectively. This points to an antiferromagnetic spin configuration, as shown schematically in (h).}
    \label{fig:CrTe2_ARPES_STM}
   \end{figure*}
   
Consistent with this, we find only modest changes in the electronic structure when heating the sample (Fig.~\ref{fig:Cr2Te3_ARPES}(c,d)), with no Stoner collapse of the exchange splitting. However, as well as an increase in the measured linewidth, we are able to resolve finite shifts in the measured binding energy of the Cr-derived states (Fig.~\ref{fig:Cr2Te3_ARPES}(d,e)). In CrGeTe$_3$~\cite{Watson2020,trzaska2023charge}, similar temperature-dependent shifts have also been observed, onsetting at the magnetic ordering temperature and serving as a marker of superexchange-type interactions~\cite{Watson2020}. Here, we find a clear anomaly in the extracted temperature-dependent valence band positions at a temperature of around 140~K (Fig.~\ref{fig:Cr2Te3_ARPES}(e)), which coincides within experimental error with the temperature ($\sim\!150$~K) at which the observed XMCD signal becomes indistinguishable from the background level (Fig.~\ref{fig:Cr2Te3_ARPES}(f)). Thus, we conclude that such temperature-dependent band shifts observed in our ARPES measurements also act as a sensitive metric of the development of long-range magnetic order in our epitaxial Cr-Te samples here.

We show in Fig.~\ref{fig:CrTe2_ARPES_STM}(a) the equivalent low-temperature electronic structure measured for our CrTe$_2$ monolayers. In contrast to Cr$_{2+\varepsilon}$Te$_3$, this system hosts a much flatter band ca.\ 1~eV below the Fermi level, which points to rather well-localised Cr-derived states. The dispersive Te-derived states lead to bands which cross the Fermi level close to 0.5~\AA$^{-1}$, consistent with tunnelling spectra from our STM measurements (Fig.~\ref{fig:CrTe2_ARPES_STM}(b)) which show a metallic sample, but in contrast to the situation for Cr$_{2+\varepsilon}$Te$_3$. As for Cr$_{2+\varepsilon}$Te$_3$, however, our ARPES measurements (Fig.~\ref{fig:CrTe2_ARPES_STM}(c,d)) show temperature-dependent band shifts of the states at $E-E_\mathrm{F}\approx-2$~eV, onsetting at a temperature of around 140~K. 

\section{Discussion}
Together with our XMCD measurements, we thus conclude that a well-developed local Cr moment persists across the measured temperature range in ML-CrTe$_2$, with the Cr-derived band being located well below the Fermi level for all temperatures; external field-driven polarisation of this moment can explain the measured XMCD signal in Fig.~\ref{fig:XMCD}(b,c). Our temperature-dependent ARPES measurements suggest that these Cr spins order below a temperature of ca.~140~K, while our XMCD measurements rule out that this order is ferromagnetic in nature. Instead, they suggest an antiferromagnetic order, as proposed in recent spin-polarised STM measurements.~\cite{xian2022spin} Consistent with this, we show atomically-resolved STM measurements from our own CrTe$_2$ samples in Fig.~\ref{fig:CrTe2_ARPES_STM}(f). A clear zig-zag pattern is observed, which gives rise to additional $(2\times1)$ magnetic Bragg peaks in Fourier transform images (blue circle in Fig.~\ref{fig:CrTe2_ARPES_STM}(g)). Consistent with prior work~\cite{xian2022spin}, we assign this additional periodicity to magnetic contrast, reflecting a zig-zag type antiferromagnetic order (Fig.~\ref{fig:CrTe2_ARPES_STM}(h)).

Our spectroscopic measurements thus point to a strong dicotomy in the electronic and magnetic properties of binary Cr-Te monolayers with control of the Cr stoichiometry. We have established ML-CrTe$_2$ as an antiferromagnetic metal with a N\'eel temperature of $\approx\!140$~K, while ML-Cr$_{2+\varepsilon}$Te$_3$ host a ferromagnetic semiconducting ground state with similar $T_\mathrm{C}\approx\!145$~K. In this respect, we conclude that the Cr-based transition-metal chalcogenides are an ideal platform for realising tunable magnetic states in the monolayer limit. Exploiting assisted nucleation and fine control of growth temperatures and fluxes in molecular-beam epitaxy growth, we have furthermore realised routes to selectively stabilise high-coverage monolayers of metastable CrTe$_2$ and Cr$_{2+\varepsilon}$Te$_3$ to enable potential exploitation of this tunability.

\section*{Methods}
\noindent {\bf Molecular beam epitaxy:} Monolayer (ML) Cr-Te alloys were synthesized on highly oriented pyrolytic graphite (HOPG) substrates utilizing a DCA R450 Molecular Beam Epitaxy (MBE) system with a base pressure of $\sim\!1\times10^{-10}$~mbar. HOPG substrates were selected due to their van der Waals structure, compatible crystal symmetry, conductive nature, and microscopically uniform surface, making them suitable for both the epitaxial growth and subsequent spectroscopic characterisation. However, the samples have small in-plane grains with random azimuthal orientation, leading to complete azimuthal averaging in our subsequent ARPES measurements.

Cr and Te were evaporated using effusion cells at temperatures of 1015–1100$^\circ$C and 425$^\circ$C, respectively, achieving a Cr-Te flux ratio of $\sim$ 1:200. The HOPG substrates were cleaved using scotch tape immediately before being loaded into the MBE load lock. Prior to the growth, they were outgassed in the growth chamber at 800°C for approximately 25 minutes. A controlled growth mechanism was achieved via the co-evaporation of small quantities of Ge ions from an e-beam evaporator. This process facilitates higher growth rates and more uniform growth by creating additional nucleation sites on the growth substrate.~\cite{advmater2024} Here, a Ge flux of 1~nA was used, as measured by a flux monitor built in to the e-beam source. In situ Reflection High-Energy Electron Diffraction (RHEED) was utilized to monitor surface quality and phase formation. After the growth, the samples were cooled in the presence of a Te flux to prevent Te vacancy formation. 

\noindent {\bf Sample characterisation:} After the growth, samples were characterised by atomic force microscopy (AFM - Park Systems NX10, operating in non-contact mode) and for some samples also by scanning tunneling microscopy (STM). The AFM was located in an Ar-filled glovebox, with the samples transferred to the glove box immediately after removing them from the MBE vacuum system. STM were performed in a home-built low temperature STM operating in ultra-high vacuum at a base temperature of $1.6\mathrm{K}$. The samples were transferred to the STM via a vacuum suitcase, without exposing their surface to non-UHV conditions. Bias voltages are applied to the sample, with the tip at virtual ground. Tunneling spectra $g(V)$ are recorded using the standard lock-in technique. Angle-resolved photoemission was performed using our lab system, with the samples transferred {\it in situ} under UHV from the MBE to the photoemission system. Measurements were conducted using a SPECS PHOIBOS 150 hemispherical electron analyzer, with a photon energy of 21.2~eV (He I$\alpha$ from a helium discharge lamp). The beam spot size was approximately 0.8 mm. 

X-ray absorption spectroscopy (XAS) and X-ray magnetic circular dichroism (XMCD) measurements were performed at the I10 beamline of Diamond Light Source, UK. Prior to removing the samples from the growth system, they were capped with $\sim$ 1 nm (estimated from the RHEED streak intensity dropping to $\sim$40\%) of tellurium deposited at room temperature, followed by deposition of a $\sim\ $2~nm amorphous selenium layer, as estimated by when the Te RHEED streaks disappear. This protects the samples from oxidation during transfer to the beamline. The caps were thin enough, however, that we could measure XAS directly through the capping layer using total-electron yield detection. The experiments utilized monochromatic synchrotron radiation with 100\% left- and right-circular polarization over the energy range of 500 to 600 eV. The magnetic field is applied along the incident beam direction. Measurements were conducted with the field applied normal to the sample, at a temperature of 10 K. Consequently, our results reflect the out-of-plane component of the magnetization.

\

\noindent\textbf{Data Availability:} The research data supporting this publication will be made openly available upon publication.

\

\section*{Acknowledgements:}
We thank Charlotte Sanders and Philip Murgatroyd for useful discussions and Martin McLaren for technical support. We gratefully acknowledge support from the Engineering and Physical Sciences Research Council (Grant Nos.~EP/X015556/1, EP/X015599/1 and EP/M023958/1) and the Leverhulme Trust (Grant No.~RL-2016-006). We thank Diamond Light Source for access to Beamline I10 (Proposal MM33239). For the purpose of open access, the authors have applied a Creative Commons Attribution (CC BY) licence to any Author Accepted Manuscript version arising. The research data supporting this publication can be accessed at [[DOI TO BE INSERTED]].

\noindent\textbf{Competing interests:} The Authors declare no Competing Financial or Non-Financial Interests.

\section*{References}

\bibliography{Reference}
\end{document}



\title{Supplementary Information: From ferromagnetic semiconductor to anti-ferromagnetic metal in epitaxial monolayers of binary Cr$_x$Te$_y$}

\author{Naina Kushwaha}
\affiliation{SUPA, School of Physics and Astronomy, University of St Andrews, North Haugh, St Andrews, KY16 9SS, United Kingdom}
\affiliation{STFC Central Laser Facility, Research Complex at Harwell, Harwell Campus, Didcot OX11 0QX, United Kingdom}
\author{Olivia Armitage}
\affiliation{SUPA, School of Physics and Astronomy, University of St Andrews, North Haugh, St Andrews, KY16 9SS, United Kingdom}
\author{Brendan Edwards}
\affiliation{SUPA, School of Physics and Astronomy, University of St Andrews, North Haugh, St Andrews, KY16 9SS, United Kingdom}

\author{Liam Trzaska}
\affiliation{SUPA, School of Physics and Astronomy, University of St Andrews, North Haugh, St Andrews, KY16 9SS, United Kingdom}

\author{Peter Bencok}
\author{Gerrit van der Laan}
\affiliation{Diamond Light Source, Harwell Science and Innovation Campus, Didcot, OX11 ODE, United Kingdom}
\author{Peter Wahl}
\email[Correspondence to: ]{wahl@st-andrews.ac.uk.}
\affiliation{SUPA, School of Physics and Astronomy, University of St Andrews, North Haugh, St Andrews, KY16 9SS, United Kingdom}
\affiliation{Physikalisches Institut, Universität Bonn, Nussallee 12, 53115 Bonn, Germany}
\author{Phil D. C. King}
\email[Correspondence to: ]{pdk6@st-andrews.ac.uk.}
\affiliation{SUPA, School of Physics and Astronomy, University of St Andrews, North Haugh, St Andrews, KY16 9SS, United Kingdom}
\author{Akhil Rajan}
\affiliation{SUPA, School of Physics and Astronomy, University of St Andrews, North Haugh, St Andrews, KY16 9SS, United Kingdom}

\date{\today}


\maketitle

\clearpage

   \begin{figure*}
   \centering
    \includegraphics[scale=0.3]{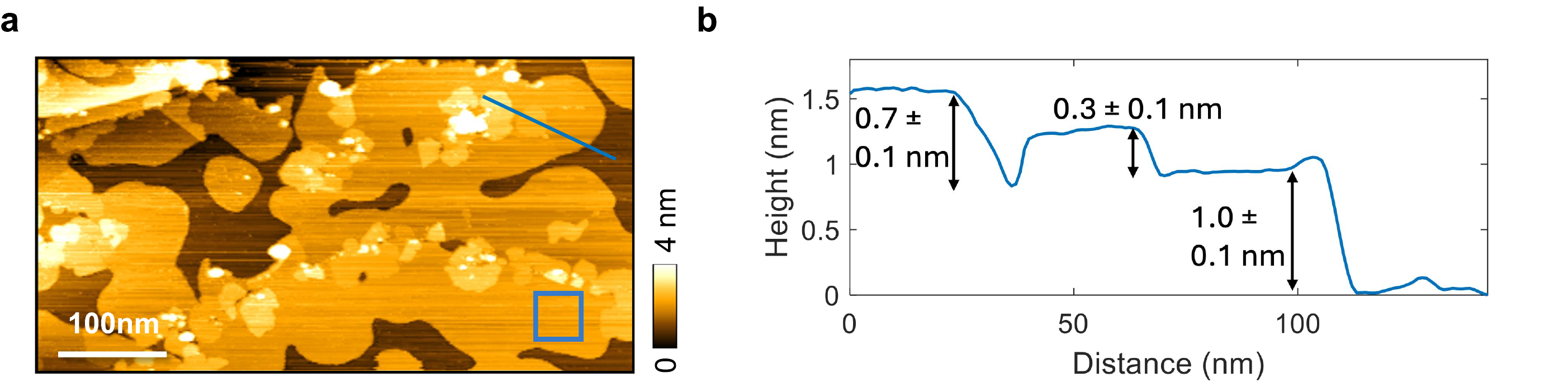}
     \caption{{\bf STM step heights.} (a) Large-area topography from STM measurements of a CrTe$_2$ sample ($V_\mathrm{s}$ = 900 mV, $I_\mathrm{s}$ = 60 pA), with occasional bilayer patches visible. (b) Height profile along the blue line in (a), showing a small step of only 0.3~nm to the next layer, reflecting formation of a Te-Cr-Te-Cr-Te region, as well as some larger step heights of 0.7~nm indicative of a self-intercalated bilayer CrTe$_2$ region. The blue box in (a) represents the region from which the atomic-resolution image shown in Fig.~5(f,g) of the main text is taken.}
   \end{figure*}

    \begin{figure*}
   \centering
    \includegraphics[scale=0.3]{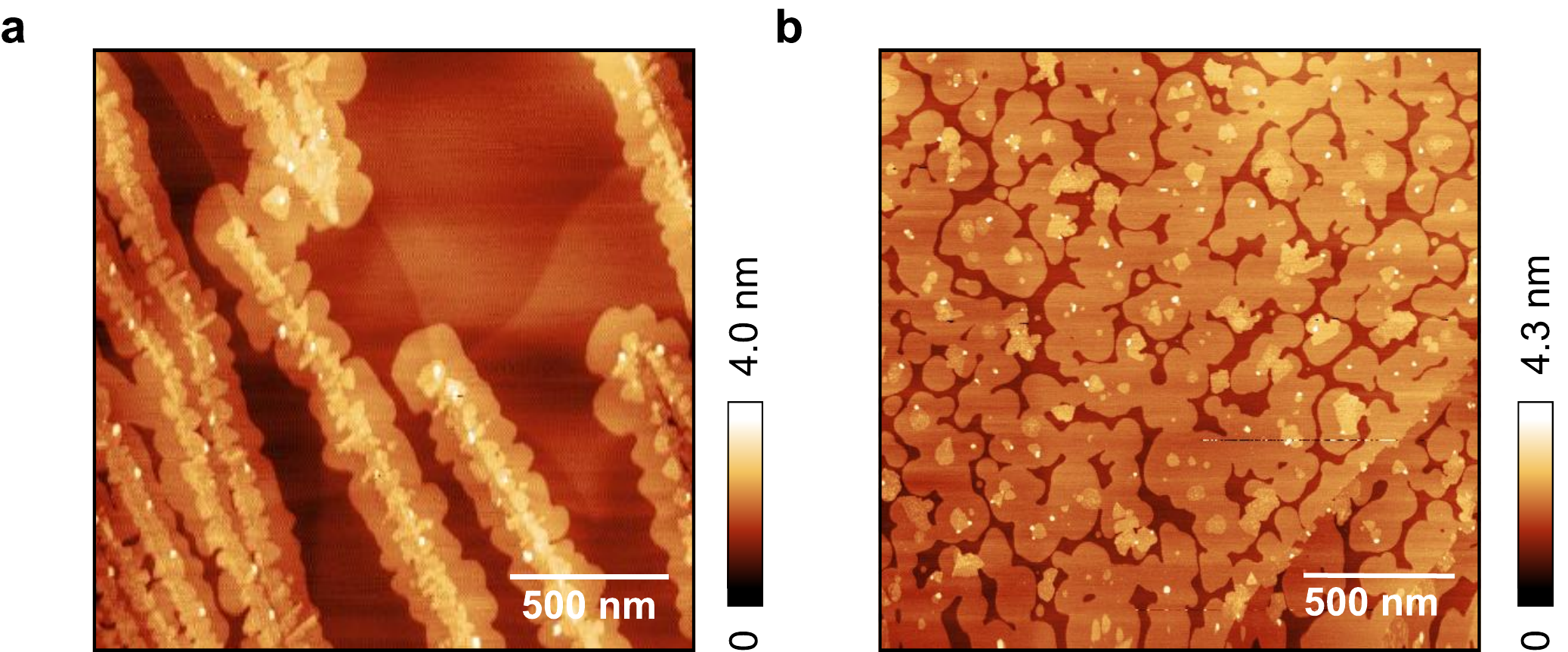}
     \caption{{\bf Effect of Ge ion co-exposure during the growth.} AFM images of CrTe$_2$ grown (a) without Ge exposure for 2 hours, and (b) with additional co-exposure to a small flux of excited Ge ions from an e-beam evaporator during a 1 hour growth. The ion-assisted growth leads to substrate defects which serve as nucleation centers for the growth. This method significantly enhances the growth rate and effectively prevents the formation of multilayer CrTe$_2$ which is higly susceptible to self-intercalation. This ion-assisted method thus results in much more stable ML 1T-CrTe$_2$ growth. }
   \end{figure*}

 \begin{figure*}
   \centering
    \includegraphics[scale=0.3]{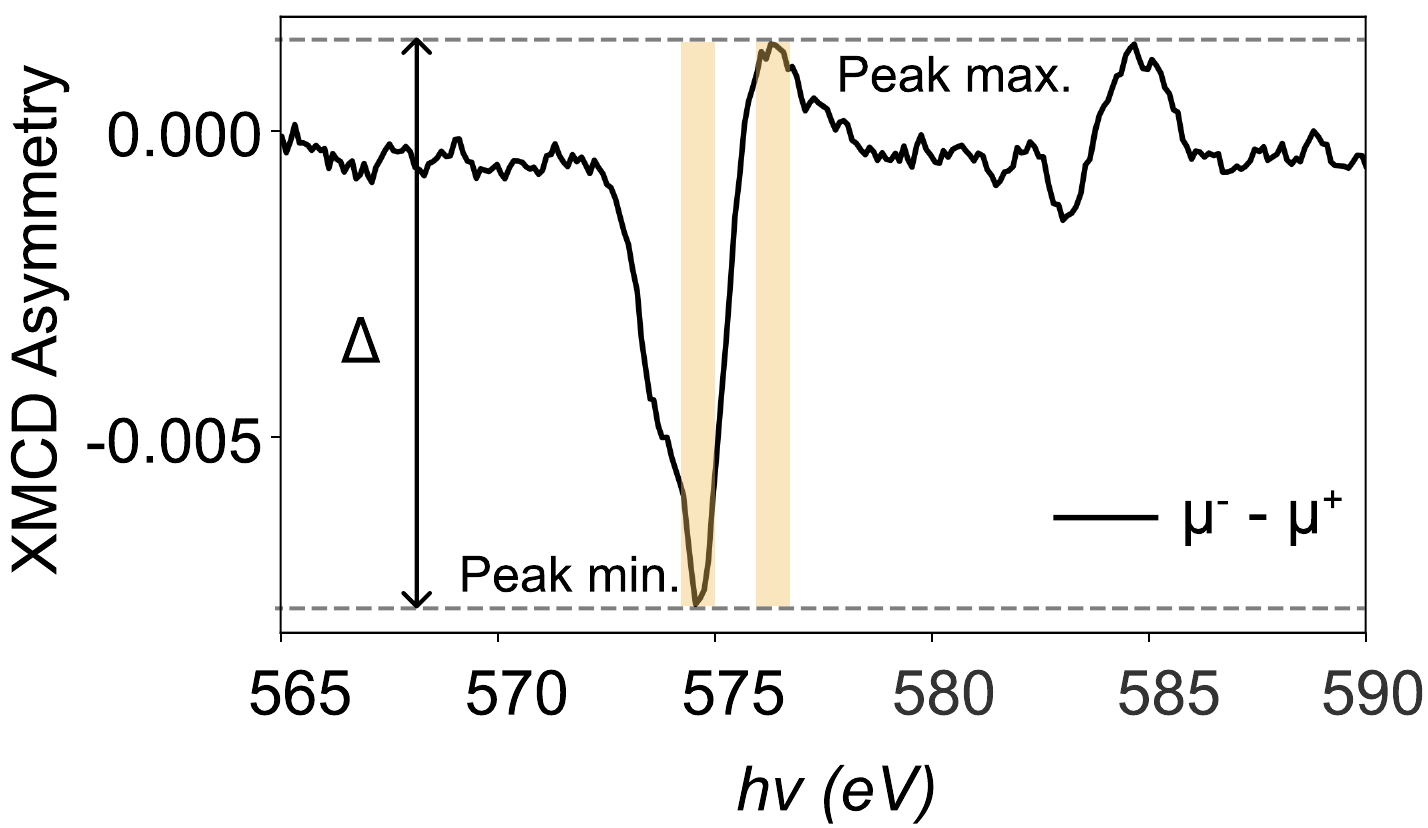}
     \caption{{\bf XMCD asymmetry measurements.} XMCD asymmetry in ML Cr$_2$Te$_3$, reproduced from Fig. 3a of the main text. The XMCD asymmetry is calculated as [$(I^{\mu^-}-I^{\mu^+})/(I^{\mu^-}+I^{\mu^+})$]. For subsequent measurements of the temperature and applied field-dependent variations, we average the asymmetry around the photon energy regions corresponding to the minimum and maximum asymmetry at the Cr $L_3$ edge (shaded regions) and take the difference in these values, $\Delta$, as a measure of the effective order parameter.}
   \end{figure*}

    \begin{figure*}
   \centering
    \includegraphics[scale=0.3]{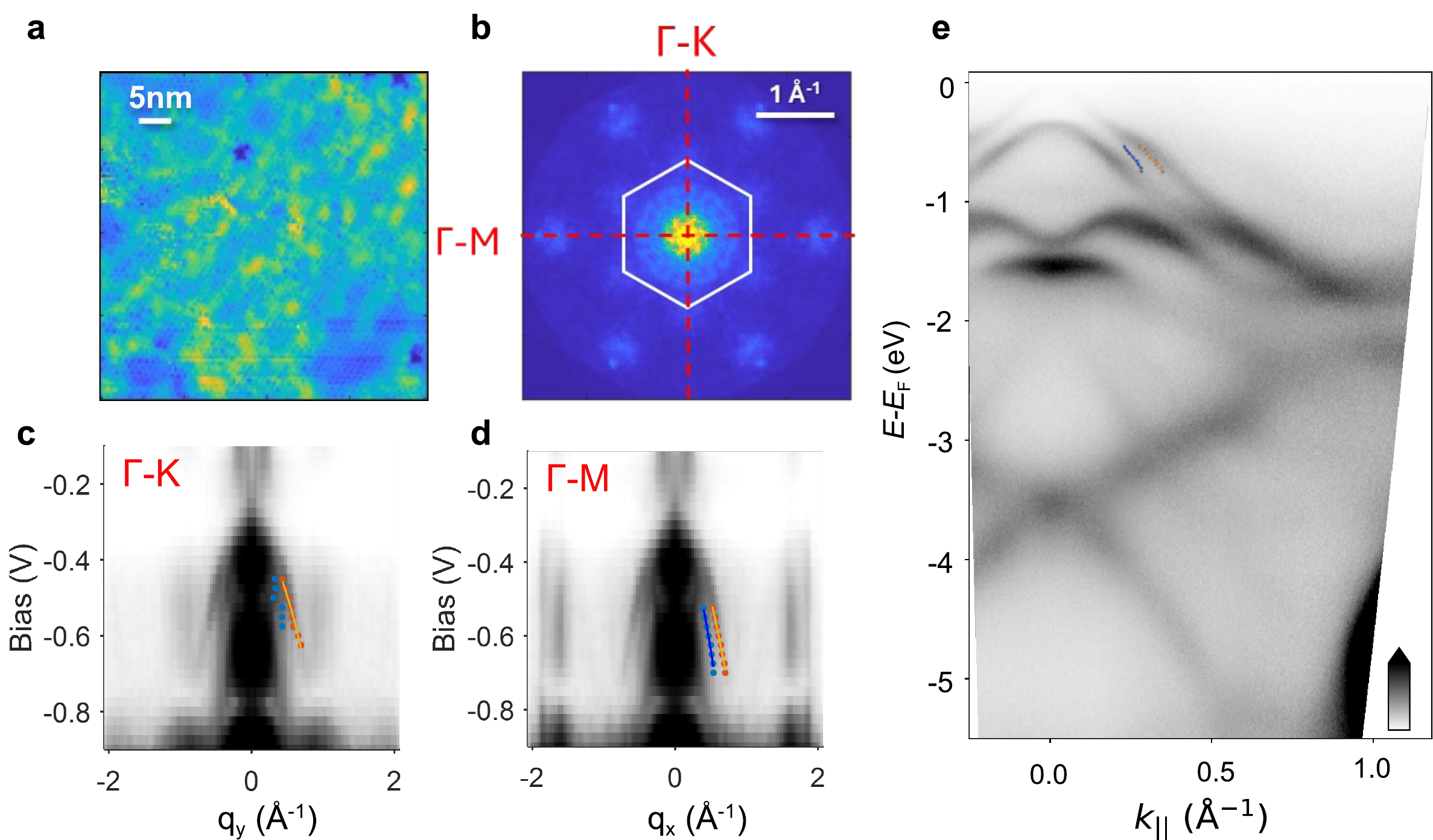}
     \caption{{\bf Quasiparticle interference in Cr$_{2+\varepsilon}$Te$_3$.} (a) - (d) Spectroscopic-imaging STM of ML Cr$_{2+\varepsilon}$Te$_3$. Spectroscopic map tunnelling setpoint $V_\mathrm{s}$ = -900 mV, $I_\mathrm{s}$ = 400 pA, lock-in amplitude 20 mV. (a) $\mathrm{d}I/\mathrm{d}V$ map layer at a bias voltage of -500 mV.  (b) Corresponding FFT. Dispersion along the (c) $\Gamma$-K, and (d) $\Gamma$-M direction from our QPI measurements. (e) Comparison of the extracted valence band dispersion with azimuthally-averaged ARPES data, showing reasonable reproduction of the observed valence band dispersions, with minor differences in the degree of hole doping.   }
   \end{figure*}